\documentclass[aps,prl,twocolumn,groupedaddress]{revtex4-1}

\usepackage[dvipdfmx]{graphicx}
\usepackage{type1cm}
\usepackage[usenames,svgnames]{xcolor}
\usepackage{color}
\usepackage{url}
\usepackage{natbib}
\usepackage[dvipdfmx]{hyperref}
\hypersetup{
     colorlinks=true,       		
     linkcolor=Navy,          	
     citecolor=Navy,            
     filecolor=Navy,      		
     urlcolor=Navy,           	
    runcolor=cyan,
 }

\usepackage{amsmath,bm}
\newcommand{\sectionprl}[1]{{\par\it #1.---}}
\def\<{\langle}
\def\>{\rangle}

\begin{document}

\title{Shortcuts to Adiabatic Classical Spin Dynamics Mimicking Quantum Annealing}

\author{Takuya Hatomura}
\email[]{hatomura@spin.phys.s.u-tokyo.ac.jp}
\affiliation{Department of Physics, The University of Tokyo, 113-8654 Tokyo, Japan}

\author{Takashi Mori}
\affiliation{Department of Physics, The University of Tokyo, 113-8654 Tokyo, Japan}


\date{\today}

\begin{abstract}
We propose a simple construction of shortcuts to adiabaticity tracking instantaneous stationary states in classical spin systems without knowing tracked stationary states. 
In our construction, control fields of counter-diabatic driving are constituted by state-dependent magnetic fields, which can be easily determined with an aid of numerical calculations. 
Easiness of our construction is a remarkable feature since it is usually a hard task to determine explicit expression of required counter-diabatic terms in many-body systems.
We also argue that our method can be applied to solve combinatorial optimization problems by considering classical spin dynamics under a time-dependent Hamiltonian, which mimics the procedure of quantum annealing. 
\end{abstract}


\maketitle

\sectionprl{Introduction}
Experimental techniques precisely tailoring quantum systems have been developed in these decades. 
They have opened up new worlds of quantum science and technology, especially quantum information processing~\cite{Nielsen2000}. 
Adiabatic control is one of the key concepts to harness quantum systems.
Adiabatic control schemes have been used for implementing adiabatic quantum computations~\cite{Farhi2000,Farhi2001}, solving combinatorial optimization problems by using quantum annealing (QA)~\cite{Kadowaki1998}, generating highly entangled states~\cite{Cirac1998,Sorensen2001,Kyaw2014}, and optimizing quantum heat engines~\cite{Allahverdyan2005}.
One of the main drawbacks of adiabatic control is long evolution time required by the adiabatic theorem~\cite{Born1928,Kato1950}, which ensures that unitary time evolution under a time-dependent Hamiltonian tracks instantaneous energy eigenstates when the Hamiltonian varies slowly enough in time. 

Theory of shortcuts to adiabaticity (STA) has been developed as a strategy to realize such adiabatic time evolution within a short time~\cite{Demirplak2003, *Demirplak2005, *Demirplak2008, Berry2009, Chen2010, delCampo2012, delCampo2013, Torrontegui2013}. 
STA enables us to realize the same time evolution without requiring slow change of a Hamiltonian by applying counter-diabatic (CD) terms instead, which are constructed by using the energy eigenstates of the original Hamiltonian~\cite{Demirplak2003, *Demirplak2005, *Demirplak2008, Berry2009}. 

It is of great interest to apply STA to above adiabatic control schemes~\cite{Farhi2000,Farhi2001,Kadowaki1998,Cirac1998,Sorensen2001,Kyaw2014,Allahverdyan2005}. 
Recently, STA has been applied to improve performance of quantum heat engines~\cite{Deng2013,delCampo2014,Beau2016}, to create highly entangled states that can be used as resources of quantum metrology and quantum computation~\cite{Julia-Diaz2012,Hatomura2018,Kyaw2018}, to accelerate primitive processes of adiabatic quantum computation~\cite{Santos2015,Santos2016}, and to speedup QA in a simple model~\cite{Takahashi2017}. 
However, application of STA to quantum many-body systems is limited due to the requirements of knowing instantaneous energy eigenstates and of implementing non-local and many-body control Hamiltonians. 
In particular, the requirement of knowing instantaneous energy eigenstates makes difficult to apply STA to QA because what we have to know is nothing but what we want to know. 

STA for classical systems has also been formulated, in which CD terms are constructed to conserve volume of phase space enclosed by equal energy surfaces, i.e., the adiabatic invariants~\cite{Jarzynski2013,Deffner2014}. 
Application of classical STA to many-body systems is also difficult because calculation of equal energy surfaces for many-body systems is hardly possible except for some special cases. 
However, the correspondence between quantum and classical STA~\cite{Okuyama2017} encourages us to investigate classical STA in detail.

In this article, we propose a simple construction of STA tracking instantaneous stationary solutions of classical spin dynamics. 
In this construction, we do not need to know tracked stationary states. 
Control fields of CD driving are given by state-dependent magnetic fields, which can be easily obtained by numerical calculations.
Our result also suggests easy implementation of STA in experiments.
Moreover, our method offers a classical algorithm for solving combinatorial optimization problems by considering classical spin dynamics mimicking the procedure of QA, which enables us to speedup each annealing process. 

Note that there is a method to construct \textit{approximate} CD terms for quantum systems without knowing instantaneous energy eigenstates based on a variational approach~\cite{Sels2017} (see also \footnote{We can construct exact CD terms if all possible operators are taken into account for trial CD terms. However, it is usually difficult for many-body systems. }). 
In contrast, our method can obtain \textit{exact} CD terms for classical systems without knowing instantaneous stationary states.

%
%

\sectionprl{Classical spin dynamics}
We consider a classical spin system consisting of $N$ spins expressed by three-dimensional unit vectors $\bm{m}_i=(m_i^x,m_i^y,m_i^z)$, $|\bm{m}_i|=1$, $i=1,2,\cdots,N$. 
Suppose that the system is described by a time-dependent Hamiltonian $\mathcal{H}_t(\{\bm{m}_i\})$.
The classical equations of motion are given by
\begin{equation}
\dot{\bm{m}}_i(t)=2\bm{m}_i(t)\times\bm{h}_i^{\mathrm{eff}}(t),
\label{eq:EOM}
\end{equation}
where $\bm{h}_i^{\mathrm{eff}}(t)$ denotes an effective field at $i$th spin, which is given by
\begin{equation}
\bm{h}_i^{\mathrm{eff}}(t)=-\frac{\partial\mathcal{H}_t}{\partial\bm{m}_i}.
\label{eq:eff}
\end{equation}
These equations of motion can be viewed as the classical limit of the Heisenberg equations $id\hat{\bm{\sigma}}_i(t)/dt=[\hat{\bm{\sigma}}_i(t),\hat{\mathcal{H}}(t)]$ under the corresponding quantum spin Hamiltonian $\hat{\mathcal{H}}(t)=\mathcal{H}_t(\{\hat{\bm{\sigma}}_i\})$, where $\hat{\bm{\sigma}}_i=(\hat{\sigma}_i^x,\hat{\sigma}_i^y,\hat{\sigma}_i^z)$ denotes the Pauli matrices describing $i$th spin. 
Here and hereafter we put $\hbar=1$. 
We can confirm that the classical equations of motion (\ref{eq:EOM}) can be also written in the form of Hamiltonian dynamics for canonical variables $\{q_i,p_i\}_{i=1}^N$ defined by
\begin{equation}
\left\{
\begin{split}
&m_i^x=\sqrt{1-(2q_i)^2}\cos p_i, \\
&m_i^y=-\sqrt{1-(2q_i)^2}\sin p_i, \\
&m_i^z=2q_i, 
\end{split}
\right.
\label{eq:canonical}
\end{equation}
i.e., a set of the equations of motion (\ref{eq:EOM}) is equivalent to that of the Hamilton equations
\begin{equation}
\dot{q}_i=\frac{\partial\mathcal{H}_t}{\partial p_i}, \quad \dot{p}_i=-\frac{\partial\mathcal{H}_t}{\partial q_i},
\end{equation}
(see, e.g., \cite{vanHemmen1986}). 

An instantaneous stationary state at time $t$, which is defined by $\{\bm{m}_i\}$ satisfying $\bm{m}_i\times\bm{h}_i^\mathrm{eff}(t)=0$ for all $i$, is specified by a minimum of $\mathcal{H}_t$ as a function of $\bm{z}=\{q_1,q_2,\dots q_N,p_1,p_2,\dots,p_N\}$, i.e.,
\begin{equation}
\frac{\partial\mathcal{H}_t}{\partial\bm{z}}=0.
\end{equation}
The instantaneous stationary state corresponding to the global minimum of $\mathcal{H}_t$ is called the instantaneous ground state.
We say that a stationary state $\{\bm{m}_i\}$ at a point $\bm{z}$ is \textit{critical} if the determinant of the Hessian matrix at this point is also zero,
\begin{equation}
\det\left[\frac{\partial^2\mathcal{H}_t}{\partial z_i\partial z_j}\right]=0, 
\label{eq:critical}
\end{equation}
i.e., there is a flat direction.

\sectionprl{Shortcuts to adiabaticity}
Now we introduce STA~\cite{Demirplak2003,*Demirplak2005,*Demirplak2008,Berry2009}. 
First we consider a generic quantum system described by a time-dependent Hamiltonian
\begin{equation}
\mathcal{\hat{H}}(t)=\sum_nE_n(t)|n(t)\rangle\langle n(t)|, 
\label{eq:qH_general}
\end{equation}
where $|n(t)\rangle$ is the energy eigenstate corresponding to the energy eigenvalue $E_n(t)$.
In STA, diabatic transitions due to time dependence of the Hamiltonian (\ref{eq:qH_general}) are canceled out by applying the following control Hamiltonian
\begin{equation}
\mathcal{\hat{H}}^\mathrm{cd}(t)=i\sum_n(1-|n(t)\rangle\langle n(t)|)|\partial_tn(t)\rangle\langle n(t)|,
\label{eq:CD}
\end{equation}
which is called the CD terms.
We can show that a solution $|\Psi(t)\rangle$ of the Schr\"odinger equation
\begin{equation}
i\frac{d}{dt}|\Psi(t)\rangle=[\hat{\mathcal{H}}(t)+\hat{\mathcal{H}}^\mathrm{cd}(t)]|\Psi(t)\rangle
\end{equation}
coincides with adiabatic dynamics under the Hamiltonian (\ref{eq:qH_general}).

As an example that is relevant to classical spin dynamics, let us consider STA for a two-level system (for derivation, see, e.g.,~\cite{Berry2009}). 
For a two-level system
\begin{equation}
\hat{\mathcal{H}}(t)=-\bm{h}(t)\cdot\hat{\bm{\sigma}},
\end{equation}
the CD Hamiltonian is given by
\begin{equation}
\hat{\mathcal{H}}^\mathrm{cd}(t)=\bm{f}(t)\cdot\hat{\bm{\sigma}},
\label{eq:CD_single}
\end{equation}
where
\begin{equation}
\bm{f}(t)=\frac{\bm{h}(t)\times\dot{\bm{h}}(t)}{2|\bm{h}(t)|^2}.
\label{eq:f_i}
\end{equation}
The total Hamiltonian $\hat{\mathcal{H}}^\mathrm{tot}(t)=\hat{\mathcal{H}}(t)+\hat{\mathcal{H}}^\mathrm{cd}(t)$ is thus given by
\begin{equation}
\hat{\mathcal{H}}^\mathrm{tot}(t)=-\left[\bm{h}(t)-\bm{f}(t)\right]\cdot\hat{\bm{\sigma}}.
\end{equation}

\sectionprl{Method}
We point out that the above CD Hamiltonian for a two-level system can be used to construct STA tracking instantaneous stationary states in classical spin systems.
This is because classical spin systems can be described by using product states of two-level systems. 
CD terms for a classical spin system with a Hamiltonian $\mathcal{H}_t$ is given by
\begin{equation}
\mathcal{H}_t^{\mathrm{cd}}=\sum_{i=1}^N\bm{f}_i(t)\cdot\bm{m}_i,
\label{eq:CD_classical}
\end{equation}
where
\begin{equation}
\bm{f}_i(t)=\frac{\bm{h}_i^\mathrm{eff}(t)\times\dot{\bm{h}}_i^\mathrm{eff}(t)}{2|\bm{h}_i^\mathrm{eff}(t)|^2}.
\label{eq:f_i_classical}
\end{equation}
This CD Hamiltonian is obtained by just replacing $\bm{h}(t)\to\bm{h}_i^\mathrm{eff}(t)$ and $\hat{\bm{\sigma}}\to\bm{m}_i$, and by taking summation over $i$ in Eqs.~(\ref{eq:CD_single}) and (\ref{eq:f_i}).
We can show that the solution of the classical equations of motion
\begin{equation}
\dot{\bm{m}}_i(t)=2\bm{m}_i(t)\times\left[\bm{h}_i^{\mathrm{eff}}(t)-\bm{f}_i(t)\right]
\label{eq:EOM_STA}
\end{equation}
tracks an instantaneous stationary state of $\mathcal{H}_t$, i.e., the solution $\{\bm{m}_i(t)\}$ satisfies $\bm{m}_i(t)\times\bm{h}_i^\mathrm{eff}(t)=0$ for all $i$, if the initial state is stationary, i.e., $\bm{m}_i(0)\times\bm{h}_i^\mathrm{eff}(0)=0$ for all $i$.
Indeed, by using Eq.~(\ref{eq:EOM_STA}), the time derivative of the following quantity
\begin{equation}
C_i(t)=\bm{m}_i(t)\cdot\frac{\bm{h}_i^\mathrm{eff}(t)}{|\bm{h}_i^\mathrm{eff}(t)|},
\end{equation}
which is identical to $\cos\theta_i(t)$, where $\theta_i(t)$ is the angle between $\bm{m}_i(t)$ and $\bm{h}_i^\mathrm{eff}(t)$, becomes
\begin{equation}
\dot{C}_i(t)=-2[\bm{m}_i(t)\times\bm{f}_i(t)]\cdot\frac{\bm{h}_i^\mathrm{eff}(t)}{|\bm{h}_i^\mathrm{eff}(t)|}+\bm{m}_i(t)\cdot\frac{d}{dt}\frac{\bm{h}_i^\mathrm{eff}(t)}{|\bm{h}_i^\mathrm{eff}(t)|}. 
\label{Eq.dotC}
\end{equation}
After a straightforward calculation, we obtain
\begin{equation}
\begin{aligned}
\bm{m}_i(t)\times\bm{f}_i(t)=&\frac{[\bm{m}_i(t)\cdot\dot{\bm{h}}_i^\mathrm{eff}(t)]\bm{h}_i^\mathrm{eff}(t)}{2|\bm{h}_i^\mathrm{eff}(t)|^2} \\
&-\frac{[\bm{m}_i(t)\cdot\bm{h}_i^\mathrm{eff}(t)]\dot{\bm{h}}_i^\mathrm{eff}(t)}{2|\bm{h}_i^\mathrm{eff}(t)|^2},
\label{eq:first_term}
\end{aligned}
\end{equation}
which is derived by using Eq.~(\ref{eq:f_i_classical}), and
\begin{equation}
\frac{d}{dt}\frac{\bm{h}_i^\mathrm{eff}(t)}{|\bm{h}_i^\mathrm{eff}(t)|}=\frac{\dot{\bm{h}}_i^\mathrm{eff}(t)}{|\bm{h}_i^\mathrm{eff}(t)|}-\frac{[\bm{h}_i^\mathrm{eff}(t)\cdot\dot{\bm{h}}_i^\mathrm{eff}(t)]\bm{h}_i^\mathrm{eff}(t)}{|\bm{h}_i^\mathrm{eff}(t)|^3}.
\label{eq:second_term}
\end{equation}
By substituting Eqs.~(\ref{eq:first_term}) and (\ref{eq:second_term}) into Eq.~(\ref{Eq.dotC}), we obtain $\dot{C}_i(t)=0$, i.e., the angle $\theta_i(t)$ between $\bm{m}_i(t)$ and $\bm{h}_i^\mathrm{eff}(t)$ is conserved. 
Therefore, the initial condition $\bm{m}_i(0)\times\bm{h}_i^\mathrm{eff}(0)=0$ leads to $\bm{m}_i(t)\times\bm{h}_i^\mathrm{eff}(t)=0$ for all time $t$, and hence we can track an instantaneous stationary state of $\mathcal{H}_t$ by applying the CD field~(\ref{eq:f_i_classical}). 
Note that STA fails when an instantaneous stationary state undergoes criticality, i.e., it satisfies Eq.~(\ref{eq:critical}), because it leads to divergence of CD fields $\bm{f}_i(t)$. 

For some applications, we want to track the instantaneous \textit{ground} state, i.e., the stationary state with the minimum energy.
In such a case, the presence of a first order transition accompanying a discontinuous jump of the ground state also matters because the ground state becomes a metastable state there.
Thus, our method succeeds in obtaining the target ground state if there is neither criticality nor a first order transition. 
It should be noted that a first order transition does not lead to divergence of CD fields in our method, whereas divergence happens in quantum STA (see, e.g., \cite{Berry2009}). 

The CD field $\bm{f}_i(t)$ depends on $\{\bm{m}_j\}$ and $\{\dot{\bm{m}}_j\}$, which is of a mean-field character.
Indeed, we can derive Eqs.~(\ref{eq:CD_classical}) and (\ref{eq:f_i_classical}) for the classical Hamiltonian $\mathcal{H}_t(\{\bm{m}_i\})$ as a result of the mean-field approximation for the corresponding quantum Hamiltonian $\hat{\mathcal{H}}(t)=\mathcal{H}_t(\{\hat{\bm{\sigma}}_i\})$~\cite{Hatomura2017}.
Because of this mean-field feature, the equations of motion~(\ref{eq:EOM_STA}) can be regarded as the self-consistent equations for $\{\dot{\bm{m}}_i\}$.
Since the set of equations~(\ref{eq:EOM_STA}) is linear in $\{\dot{\bm{m}}_i\}$, it is not hard to solve these self-consistent equations.

In this way, we can easily perform STA in classical spin systems by just applying additional magnetic fields~(\ref{eq:f_i_classical}). 
It is in stark contrast to quantum many-body systems, in which it is in general a hard task to obtain explicit expression of CD terms since it depends on energy eigenstates of many-body Hamiltonians (see Eq.~(\ref{eq:CD})).
Even if we could obtain CD terms in quantum many-body systems, it would also be very hard to implement in experiments because CD terms contain non-local and many-body interactions.

\begin{figure}
\includegraphics[width=8cm]{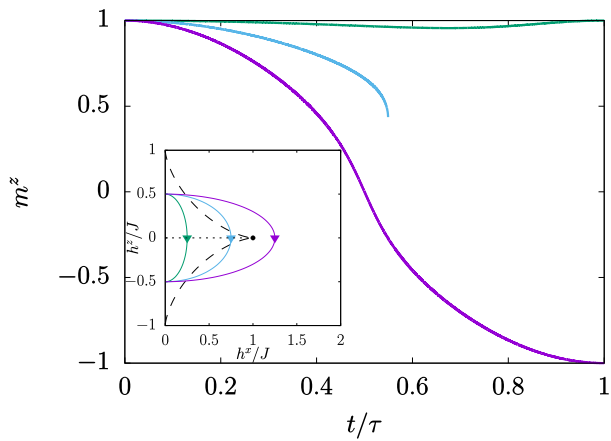}
\caption{\label{Fig.LMG} (Color online) Stationary magnetization dynamics tracked by STA. Our method results in (i; purple) the exact ground state, (ii; green) the metastable state due to a first order transition, and (iii; cyan) divergence due to criticality. (Inset) Paths in parameter space and the phase diagram. }
\end{figure}

\sectionprl{Demonstration in a simple model}
Now we demonstrate our method by using the following paradigmatic model
\begin{equation}
\mathcal{H}_t=-\frac{J}{2N}\sum_{i,j=1}^Nm_i^zm_j^z-h^z(t)\sum_{i=1}^Nm_i^z-h^x(t)\sum_{i=1}^Nm_i^x, 
\label{Eq.LMG}
\end{equation}
where the coupling strength $J$ is a positive constant. 
We show when our method can find the exact ground state and how transitions and criticality affect stationary state tracking by STA. 
In this model, first order transitions take place in the ground state when the parameters cross the transition line $h^z(t)=0$ and $h^x(t)\in(-J,J)$, which is represented by a dotted line in the inset of Fig.~\ref{Fig.LMG}.
Inside the spinodal lines $J^{2/3}=(h^z(t))^{2/3}+(h^x(t))^{2/3}$ (dashed lines in the inset of Fig.~\ref{Fig.LMG}), there are two stationary states, i.e., the ground state and the metastable state, while the stationary state is unique in the outside region.
The ground state shows criticality at the point specified by $J=|h^x(t)|$ and $h^z(t)=0$ (a black point in the inset of Fig.~\ref{Fig.LMG}), while the metastable state shows criticality at the spinodal lines.

We simulate the following three cases: (i) no transition takes place, (ii) a first order transition takes place, and (iii) the system undergoes criticality after a first order transition. 
As mentioned above, our method will result in (i) the exact ground state, (ii) the metastable state, and (iii) divergence, respectively. 
We assume that the magnetic fields are given by $h^z(t)=J\cos[\pi t/\tau]/2$ and $h^x(t)=h_0\sin[\pi t/\tau]$, where $h_0\ (>0)$ enables us to change the path in parameter space and $\tau$ is the operation time. 
In this setup, we can test above three cases with the same initial Hamiltonian and with the same final Hamiltonian. 
That is the parameters of the initial Hamiltonian are $(h^x(0),h^z(0))=(0,J/2)$ and those of the final Hamiltonian are $(h^x(\tau),h^z(\tau))=(0,-J/2)$, and the system undergoes a first order transition when $0<h_0/J<1$ and shows criticality when $1/2\le h_0/J\le1$. 
Note that the ground state of the final Hamiltonian is given by the all spin-down state and the metastable state of that is given by the all spin-up state. 
We perform numerical simulations with the parameters (i) $h_0/J=5/4$ (purple curves), (ii) $h_0/J=1/4$ (green curves), and (iii) $h_0/J=3/4$ (cyan curves), and depict $m^z(t)\equiv\sum_{i=1}^Nm_i^z(t)/N$ in Fig.~\ref{Fig.LMG}. 
The result of numerical simulations clearly shows properties of our method. 
Note that the results in Fig.~\ref{Fig.LMG} do not depend on $N$, $\tau$, and $J$.

\sectionprl{Mimicking the procedure of quantum annealing}
Next we consider to solve combinatorial optimization problems, which can be formulated as a problem to find the ground state of the Ising Hamiltonian
\begin{equation}
\hat{\mathcal{H}}_T=-\frac{1}{2}\sum_{i,j=1}^NJ_{ij}\hat{\sigma}_i^z\hat{\sigma}_j^z-\sum_{i=1}^Nh_i^z\hat{\sigma}_i^z,
\label{eq:target}
\end{equation}
called the target Hamiltonian. 
In QA, we utilize the transverse field Hamiltonian $\hat{\mathcal{V}}=-\sum_{i=1}^N\hat{\sigma}_i^x$ as a source of quantum fluctuations to find the ground state of the target Hamiltonian $\hat{\mathcal{H}}_T$~\cite{Tanaka-Horiguchi-Jpn, *Tanaka2000, Kadowaki1998, Brooke1999, Farhi2000, Farhi2001, Santoro2427, Martonak2002}. 
We change the Hamiltonian as
\begin{equation}
\hat{\mathcal{H}}(t)=g(t/\tau)\hat{\mathcal{H}}_T+[1-g(t/\tau)]\hat{\mathcal{V}},
\label{eq:QA}
\end{equation}
where $g(t/\tau)$ is a continuous function of time satisfying $g(0)=0$ and $g(1)=1$, and $\tau$ is the annealing time. 
In this article, we assume $g(t/\tau)=[1-\cos(\pi t/\tau)]/2$, which satisfies $g^\prime(0)=g^\prime(1)=0$ and thus CD fields vanish at the initial and final time. 
The initial state is prepared as the ground state of the initial Hamiltonian $\hat{\mathcal{H}}(0)=\hat{\mathcal{V}}$, and then the adiabatic theorem ensures that the system remains in the instantaneous ground state and finally reaches the ground state of the final Hamiltonian $\hat{\mathcal{H}}(\tau)=\hat{\mathcal{H}}_T$ if the annealing time $\tau$ is sufficiently large. 
It means that we can solve combinatorial optimization problems. 

Nowadays, we can implement QA by using quantum annealers, e.g., the D-Wave machine~\cite{Johnson2011}. 
Recent argument about quantumness of the D-Wave machine yielded new algorithms using \textit{classical} spin dynamics that mimics the methodology of QA~\cite{Smolin2014, Wang2013, Boixo2014, Shin2014, Albash2015reexamining}. 
Strong correlations between performance of the D-Wave machine and that of those classical algorithms have been reported~\cite{Smolin2014,Shin2014,Albash2015reexamining}. 
Although it is still under discussion if those classical algorithms can simulate true QA, they are at least useful to obtain approximate solutions of combinatorial optimization problems. 

Now we consider a classical analog of the QA Hamiltonian~(\ref{eq:QA}) expressed by the time-dependent classical Hamiltonian
\begin{equation}
\mathcal{H}_t=g(t/\tau)\mathcal{H}_T+[1-g(t/\tau)]\mathcal{V},
\label{eq:CA}
\end{equation}
where $\mathcal{H}_T$ is the classical target Hamiltonian
\begin{equation}
\mathcal{H}_T=-\frac{1}{2}\sum_{i,j=1}^NJ_{ij}m_i^zm_j^z-\sum_{i=1}^Nh_i^zm_i^z,
\label{eq:target_classical}
\end{equation}
and $\mathcal{V}$ is the classical transverse field Hamiltonian $\mathcal{V}=-\sum_{i=1}^Nm_i^x$. 
Starting from the ground state of the initial Hamiltonian $\mathcal{H}_0=\mathcal{V}$, we expect to reach the ground state of the final Hamiltonian $\mathcal{H}_{\tau}=\mathcal{H}_T$ in the limit of $\tau\to\infty$ or by using our method. 
However, as demonstrated by using the model (\ref{Eq.LMG}), classical algorithms relying on deterministic classical dynamics result in failure to obtain the exact ground state when a stationary state undergoes transitions and/or criticality. 
It is known that first order transitions are sometimes resolved when we apply inhomogeneous driving~\cite{Rams2016,Susa2018}.
Here, we consider the random transverse-field Hamiltonian
\begin{equation}
\mathcal{V}'=-\sum_{i=1}^Nh_i^xm_i^x,
\end{equation}
instead of $\mathcal{V}$ in Eq.~(\ref{eq:CA}).

We numerically test our method by using the random field Ising model on the $L\times L$ square lattice, i.e., $J_{ij}=1$ for the neighboring pairs and $J_{ij}=0$ otherwise, and $\{h_i^z\}$ are random variables taking $h_i^z=\pm 0.3$. 
The number of spins is given by $N=L^2$. 
The ground state of this model can be exactly obtained by using the max-flow-min-cut algorithm (see, e.g., \cite{Rieger1998}). 
For a given realization of $\{h_i^z\}$, we perform our method mimicking the annealing procedure with $\tau=1$ by solving Eq.~(\ref{eq:EOM_STA}) for $M$ realizations of the random transverse fields $\{h_i^x\}$.
The minimum energy among these $M$ realizations, which is denoted by $E_{\mathrm{est}}$, is compared to the exact ground state energy $E_g$ obtained by using the max-flow-min-cut algorithm.
The result is regarded as a failure if the difference of the energies measured by $\Delta=|(E_{\mathrm{est}}-E_g)/E_g|$ is greater than 0.01.
Note that the total computation time is proportional to $M$ since we repeat the annealing procedure $M$ times under different realizations of $\{h_i^x\}$.

\begin{figure}
\centering
\includegraphics[width=8cm]{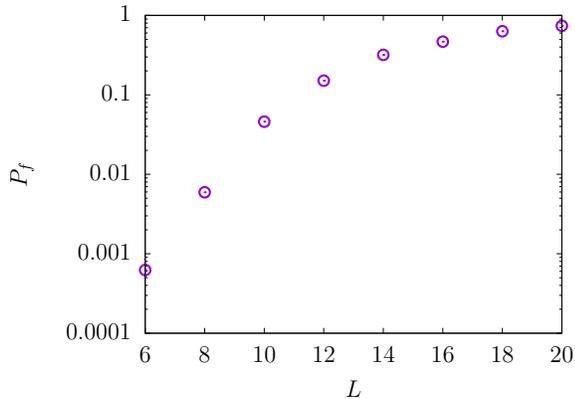}
\caption{\label{Fig.uniform}Failure probability with respect to the system size. The number of spins is given by $N=L^2$. }
\end{figure}

In Fig.~\ref{Fig.uniform}, the system-size dependence of the failure probability for 3,456 realizations of $\{h_i^z\}$ is plotted in the case of the uniform transverse field ($h_i^x=1$ for all $i$ and $M=1$).
For small system sizes $L\leq 8$ ($N\le64$), the failure probability is less than $1\%$, but it grows rapidly as the system size increases.
The failure is due to the occurrence of first order transitions or criticality in the ground state.
This result implies that solving combinatorial optimization problems by using classical models of QA with a uniform transverse field is intrinsically difficult and it might also imply difficulty to solve them by using QA with a uniform transverse field because of classical and quantum correspondence.

\begin{figure}
\centering
\includegraphics[width=8cm]{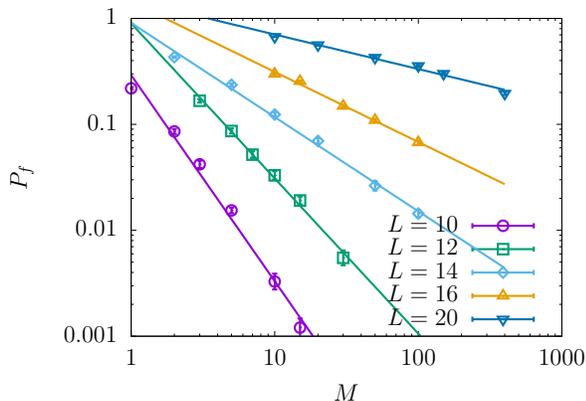}
\caption{\label{Fig.inhomo}(Color online) Failure probability with respect to the repetition of inhomogeneous driving. The system size is depicted from $L=10$ to $L=20$ (from $N=100$ to $N=400$). The error bars represent the standard errors of the binomial distribution. }
\end{figure}

\begin{figure}
\centering
\includegraphics[width=8cm]{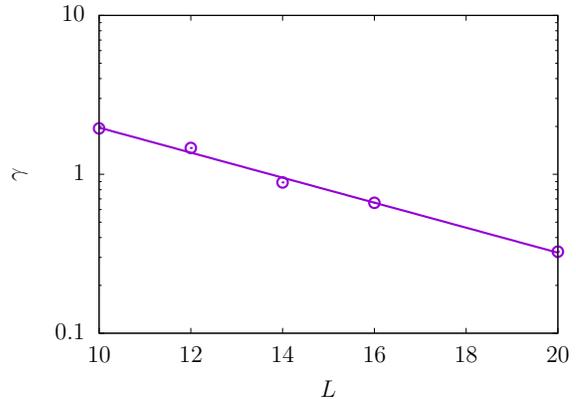}
\caption{\label{Fig.slope}Decreasing rate of the failure probability with respect to the system size. }
\end{figure}

Next, we repeat $M$ different realizations of the random transverse fields $\{h_i^x\}$ for each realization of $\{h_i^z\}$ in order to avoid the above difficulty due to first order transitions and criticality. 
We choose $h_i^x$ uniformly from the interval $h_i^x\in[1,2]$ for each $i$. 
The repetition dependence of the failure probability is depicted in Fig.~\ref{Fig.inhomo}. 
The plot shows that for a fixed system size $N$ the failure probability asymptotically decreases as $P_f\sim M^{-\gamma}$ with an exponent $\gamma$.
Thus, we can avoid the occurrence of transitions and criticality by increasing $M$. 
However, as shown in Fig.~\ref{Fig.slope}, the exponent $\gamma$ decreases in the exponential way $\gamma\sim e^{-\mathcal{O}(N)}$, and thus inhomogeneous driving based on uniform random numbers $\{h_i^x\}$ is not so efficient for large system sizes even if STA is applied. 
Note that some least squares fitting for small $M$ in Fig.~\ref{Fig.inhomo} tends to above unity. 
This is because the above decreasing rates of the failure probability are asymptotic behavior and those for small $M$ are much slow.

\sectionprl{Conclusion}
In this article, we proposed a simple construction of STA for classical spin systems tracking instantaneous stationary states without knowing tracked instantaneous stationary states. 
In contrast, in order to construct CD terms, energy eigenstates are required in quantum cases and volume of phase space is required in classical cases in previous works. 
Starting from a stationary state of an initial Hamiltonian, our method results in one of the stationary states of a final Hamiltonian with arbitrary time scale if there is no criticality. 
Although we have applied our method to simple Hamiltonians with two-body Ising interactions and local magnetic fields in this article, our method is applicable to arbitrary classical spin Hamiltonians $\mathcal{H}_t(\{\bm{m}_i\})$.

Our method can be used to solve combinatorial optimization problems by mimicking the procedure of QA, where we can speedup each annealing process. 
In this algorithm, we aim to track the instantaneous ground state of a classical spin system within a short time.
In this case, not only criticality but also first order transitions matter because the ground state becomes a metastable state at a first order transition point. 
As a demonstration, we applied our method to the random field Ising model on the square lattice. 
From our observations, we could suggest that (i) solving combinatorial optimization problems by using QA with a uniform transverse field would be difficult because criticality and first order transitions would rapidly increase along with system size even in quantum cases and (ii) inhomogeneous driving in QA for random systems could also resolve first order transitions but (iii) it is not so efficient if we use simple inhomogeneity such as i.i.d. random transverse fields. 
For these difficulties (i) and (iii), for examples, (i) QA using non-stoquastic Hamiltonians and (iii) somehow designed inhomogeneous driving could be candidates for resolution, respectively. 
These are left for future works. 

\begin{acknowledgments}
\sectionprl{Acknowledgments}
The authors are grateful to Jun Takahashi for helpful comments. 
Numerical calculations are supported by the Supercomputer Center in the Institute for Solid State Physics (ISSP) of the University of Tokyo. 
The authors are supported by the Ministry of Education, Culture, Sports, Science and Technology (MEXT) of Japan through the Elements Strategy Initiative Center for Magnetic Materials (ESICMM). 
TH is supported by the Japan Society for the Promotion of Science (JSPS) through the Research Fellowship for Young Scientists (DC2) and through the Program for Leading Graduate Schools: Material Education program for the future leaders in Research, Industry, and Technology (MERIT) of the University of Tokyo. 
This work is partly supported by JSPS KAKENHI Grant Number JP18J11053. 
\end{acknowledgments}

\bibliography{QAviaSTAbib}

\begin{thebibliography}{48}%
\makeatletter
\providecommand \@ifxundefined [1]{%
 \@ifx{#1\undefined}
}%
\providecommand \@ifnum [1]{%
 \ifnum #1\expandafter \@firstoftwo
 \else \expandafter \@secondoftwo
 \fi
}%
\providecommand \@ifx [1]{%
 \ifx #1\expandafter \@firstoftwo
 \else \expandafter \@secondoftwo
 \fi
}%
\providecommand \natexlab [1]{#1}%
\providecommand \enquote  [1]{``#1''}%
\providecommand \bibnamefont  [1]{#1}%
\providecommand \bibfnamefont [1]{#1}%
\providecommand \citenamefont [1]{#1}%
\providecommand \href@noop [0]{\@secondoftwo}%
\providecommand \href [0]{\begingroup \@sanitize@url \@href}%
\providecommand \@href[1]{\@@startlink{#1}\@@href}%
\providecommand \@@href[1]{\endgroup#1\@@endlink}%
\providecommand \@sanitize@url [0]{\catcode `\\12\catcode `\$12\catcode
  `\&12\catcode `\#12\catcode `\^12\catcode `\_12\catcode `\%12\relax}%
\providecommand \@@startlink[1]{}%
\providecommand \@@endlink[0]{}%
\providecommand \url  [0]{\begingroup\@sanitize@url \@url }%
\providecommand \@url [1]{\endgroup\@href {#1}{\urlprefix }}%
\providecommand \urlprefix  [0]{URL }%
\providecommand \Eprint [0]{\href }%
\providecommand \doibase [0]{http://dx.doi.org/}%
\providecommand \selectlanguage [0]{\@gobble}%
\providecommand \bibinfo  [0]{\@secondoftwo}%
\providecommand \bibfield  [0]{\@secondoftwo}%
\providecommand \translation [1]{[#1]}%
\providecommand \BibitemOpen [0]{}%
\providecommand \bibitemStop [0]{}%
\providecommand \bibitemNoStop [0]{.\EOS\space}%
\providecommand \EOS [0]{\spacefactor3000\relax}%
\providecommand \BibitemShut  [1]{\csname bibitem#1\endcsname}%
\let\auto@bib@innerbib\@empty
\bibitem [{\citenamefont {Nielsen}\ and\ \citenamefont
  {Chuang}(2000)}]{Nielsen2000}%
  \BibitemOpen
  \bibfield  {author} {\bibinfo {author} {\bibfnamefont {M.~A.}\ \bibnamefont
  {Nielsen}}\ and\ \bibinfo {author} {\bibfnamefont {I.~L.}\ \bibnamefont
  {Chuang}},\ }\href {https://books.google.co.jp/books?id=65FqEKQOfP8C} {\emph
  {\bibinfo {title} {Quantum Computation and Quantum Information}}},\ Cambridge
  Series on Information and the Natural Sciences\ (\bibinfo  {publisher}
  {Cambridge University Press},\ \bibinfo {year} {2000})\BibitemShut {NoStop}%
\bibitem [{\citenamefont {Farhi}\ \emph {et~al.}(2000)\citenamefont {Farhi},
  \citenamefont {Goldstone}, \citenamefont {Gutmann},\ and\ \citenamefont
  {Sipser}}]{Farhi2000}%
  \BibitemOpen
  \bibfield  {author} {\bibinfo {author} {\bibfnamefont {E.}~\bibnamefont
  {Farhi}}, \bibinfo {author} {\bibfnamefont {J.}~\bibnamefont {Goldstone}},
  \bibinfo {author} {\bibfnamefont {S.}~\bibnamefont {Gutmann}}, \ and\
  \bibinfo {author} {\bibfnamefont {M.}~\bibnamefont {Sipser}},\ }\href
  {https://arxiv.org/abs/quant-ph/0001106} {\bibfield  {journal} {\bibinfo
  {journal} {arXiv preprint quant-ph/0001106}\ } (\bibinfo {year}
  {2000})}\BibitemShut {NoStop}%
\bibitem [{\citenamefont {Farhi}\ \emph {et~al.}(2001)\citenamefont {Farhi},
  \citenamefont {Goldstone}, \citenamefont {Gutmann}, \citenamefont {Lapan},
  \citenamefont {Lundgren},\ and\ \citenamefont {Preda}}]{Farhi2001}%
  \BibitemOpen
  \bibfield  {author} {\bibinfo {author} {\bibfnamefont {E.}~\bibnamefont
  {Farhi}}, \bibinfo {author} {\bibfnamefont {J.}~\bibnamefont {Goldstone}},
  \bibinfo {author} {\bibfnamefont {S.}~\bibnamefont {Gutmann}}, \bibinfo
  {author} {\bibfnamefont {J.}~\bibnamefont {Lapan}}, \bibinfo {author}
  {\bibfnamefont {A.}~\bibnamefont {Lundgren}}, \ and\ \bibinfo {author}
  {\bibfnamefont {D.}~\bibnamefont {Preda}},\ }\href {\doibase
  10.1126/science.1057726} {\bibfield  {journal} {\bibinfo  {journal}
  {Science}\ }\textbf {\bibinfo {volume} {292}},\ \bibinfo {pages} {472}
  (\bibinfo {year} {2001})}\BibitemShut {NoStop}%
\bibitem [{\citenamefont {Kadowaki}\ and\ \citenamefont
  {Nishimori}(1998)}]{Kadowaki1998}%
  \BibitemOpen
  \bibfield  {author} {\bibinfo {author} {\bibfnamefont {T.}~\bibnamefont
  {Kadowaki}}\ and\ \bibinfo {author} {\bibfnamefont {H.}~\bibnamefont
  {Nishimori}},\ }\href {\doibase 10.1103/PhysRevE.58.5355} {\bibfield
  {journal} {\bibinfo  {journal} {Phys. Rev. E}\ }\textbf {\bibinfo {volume}
  {58}},\ \bibinfo {pages} {5355} (\bibinfo {year} {1998})}\BibitemShut
  {NoStop}%
\bibitem [{\citenamefont {Cirac}\ \emph {et~al.}(1998)\citenamefont {Cirac},
  \citenamefont {Lewenstein}, \citenamefont {M\o{}lmer},\ and\ \citenamefont
  {Zoller}}]{Cirac1998}%
  \BibitemOpen
  \bibfield  {author} {\bibinfo {author} {\bibfnamefont {J.~I.}\ \bibnamefont
  {Cirac}}, \bibinfo {author} {\bibfnamefont {M.}~\bibnamefont {Lewenstein}},
  \bibinfo {author} {\bibfnamefont {K.}~\bibnamefont {M\o{}lmer}}, \ and\
  \bibinfo {author} {\bibfnamefont {P.}~\bibnamefont {Zoller}},\ }\href
  {\doibase 10.1103/PhysRevA.57.1208} {\bibfield  {journal} {\bibinfo
  {journal} {Phys. Rev. A}\ }\textbf {\bibinfo {volume} {57}},\ \bibinfo
  {pages} {1208} (\bibinfo {year} {1998})}\BibitemShut {NoStop}%
\bibitem [{\citenamefont {S\o{}rensen}\ and\ \citenamefont
  {M\o{}lmer}(2001)}]{Sorensen2001}%
  \BibitemOpen
  \bibfield  {author} {\bibinfo {author} {\bibfnamefont {A.~S.}\ \bibnamefont
  {S\o{}rensen}}\ and\ \bibinfo {author} {\bibfnamefont {K.}~\bibnamefont
  {M\o{}lmer}},\ }\href {\doibase 10.1103/PhysRevLett.86.4431} {\bibfield
  {journal} {\bibinfo  {journal} {Phys. Rev. Lett.}\ }\textbf {\bibinfo
  {volume} {86}},\ \bibinfo {pages} {4431} (\bibinfo {year}
  {2001})}\BibitemShut {NoStop}%
\bibitem [{\citenamefont {Kyaw}\ \emph {et~al.}(2014)\citenamefont {Kyaw},
  \citenamefont {Li},\ and\ \citenamefont {Kwek}}]{Kyaw2014}%
  \BibitemOpen
  \bibfield  {author} {\bibinfo {author} {\bibfnamefont {T.~H.}\ \bibnamefont
  {Kyaw}}, \bibinfo {author} {\bibfnamefont {Y.}~\bibnamefont {Li}}, \ and\
  \bibinfo {author} {\bibfnamefont {L.-C.}\ \bibnamefont {Kwek}},\ }\href
  {\doibase 10.1103/PhysRevLett.113.180501} {\bibfield  {journal} {\bibinfo
  {journal} {Phys. Rev. Lett.}\ }\textbf {\bibinfo {volume} {113}},\ \bibinfo
  {pages} {180501} (\bibinfo {year} {2014})}\BibitemShut {NoStop}%
\bibitem [{\citenamefont {Allahverdyan}\ and\ \citenamefont
  {Nieuwenhuizen}(2005)}]{Allahverdyan2005}%
  \BibitemOpen
  \bibfield  {author} {\bibinfo {author} {\bibfnamefont {A.~E.}\ \bibnamefont
  {Allahverdyan}}\ and\ \bibinfo {author} {\bibfnamefont {T.~M.}\ \bibnamefont
  {Nieuwenhuizen}},\ }\href {\doibase 10.1103/PhysRevE.71.046107} {\bibfield
  {journal} {\bibinfo  {journal} {Phys. Rev. E}\ }\textbf {\bibinfo {volume}
  {71}},\ \bibinfo {pages} {046107} (\bibinfo {year} {2005})}\BibitemShut
  {NoStop}%
\bibitem [{\citenamefont {Born}\ and\ \citenamefont {Fock}(1928)}]{Born1928}%
  \BibitemOpen
  \bibfield  {author} {\bibinfo {author} {\bibfnamefont {M.}~\bibnamefont
  {Born}}\ and\ \bibinfo {author} {\bibfnamefont {V.}~\bibnamefont {Fock}},\
  }\href {\doibase 10.1007/BF01343193} {\bibfield  {journal} {\bibinfo
  {journal} {Z. Phys.}\ }\textbf {\bibinfo {volume} {51}},\ \bibinfo {pages}
  {165} (\bibinfo {year} {1928})}\BibitemShut {NoStop}%
\bibitem [{\citenamefont {Kato}(1950)}]{Kato1950}%
  \BibitemOpen
  \bibfield  {author} {\bibinfo {author} {\bibfnamefont {T.}~\bibnamefont
  {Kato}},\ }\href {\doibase 10.1143/JPSJ.5.435} {\bibfield  {journal}
  {\bibinfo  {journal} {J. Phys. Soc. Jpn.}\ }\textbf {\bibinfo {volume} {5}},\
  \bibinfo {pages} {435} (\bibinfo {year} {1950})}\BibitemShut {NoStop}%
\bibitem [{\citenamefont {Demirplak}\ and\ \citenamefont
  {Rice}(2003)}]{Demirplak2003}%
  \BibitemOpen
  \bibfield  {author} {\bibinfo {author} {\bibfnamefont {M.}~\bibnamefont
  {Demirplak}}\ and\ \bibinfo {author} {\bibfnamefont {S.~A.}\ \bibnamefont
  {Rice}},\ }\href {\doibase 10.1021/jp030708a} {\bibfield  {journal} {\bibinfo
   {journal} {J. Phys. Chem. A}\ }\textbf {\bibinfo {volume} {107}},\ \bibinfo
  {pages} {9937} (\bibinfo {year} {2003})}\BibitemShut {NoStop}%
\bibitem [{\citenamefont {Demirplak}\ and\ \citenamefont
  {Rice}(2005)}]{Demirplak2005}%
  \BibitemOpen
  \bibfield  {author} {\bibinfo {author} {\bibfnamefont {M.}~\bibnamefont
  {Demirplak}}\ and\ \bibinfo {author} {\bibfnamefont {S.~A.}\ \bibnamefont
  {Rice}},\ }\href {\doibase 10.1021/jp040647w} {\bibfield  {journal} {\bibinfo
   {journal} {J. Phys. Chem. B}\ }\textbf {\bibinfo {volume} {109}},\ \bibinfo
  {pages} {6838} (\bibinfo {year} {2005})}\BibitemShut {NoStop}%
\bibitem [{\citenamefont {Demirplak}\ and\ \citenamefont
  {Rice}(2008)}]{Demirplak2008}%
  \BibitemOpen
  \bibfield  {author} {\bibinfo {author} {\bibfnamefont {M.}~\bibnamefont
  {Demirplak}}\ and\ \bibinfo {author} {\bibfnamefont {S.~A.}\ \bibnamefont
  {Rice}},\ }\href {\doibase 10.1063/1.2992152} {\bibfield  {journal} {\bibinfo
   {journal} {J. Chem. Phys.}\ }\textbf {\bibinfo {volume} {129}},\ \bibinfo
  {pages} {154111} (\bibinfo {year} {2008})}\BibitemShut {NoStop}%
\bibitem [{\citenamefont {Berry}(2009)}]{Berry2009}%
  \BibitemOpen
  \bibfield  {author} {\bibinfo {author} {\bibfnamefont {M.~V.}\ \bibnamefont
  {Berry}},\ }\href {\doibase 10.1088/1751-8113/42/36/365303} {\bibfield
  {journal} {\bibinfo  {journal} {J. Phys. A: Math. Theor.}\ }\textbf {\bibinfo
  {volume} {42}},\ \bibinfo {pages} {365303} (\bibinfo {year}
  {2009})}\BibitemShut {NoStop}%
\bibitem [{\citenamefont {Chen}\ \emph {et~al.}(2010)\citenamefont {Chen},
  \citenamefont {Ruschhaupt}, \citenamefont {Schmidt}, \citenamefont {del
  Campo}, \citenamefont {Gu\'ery-Odelin},\ and\ \citenamefont
  {Muga}}]{Chen2010}%
  \BibitemOpen
  \bibfield  {author} {\bibinfo {author} {\bibfnamefont {X.}~\bibnamefont
  {Chen}}, \bibinfo {author} {\bibfnamefont {A.}~\bibnamefont {Ruschhaupt}},
  \bibinfo {author} {\bibfnamefont {S.}~\bibnamefont {Schmidt}}, \bibinfo
  {author} {\bibfnamefont {A.}~\bibnamefont {del Campo}}, \bibinfo {author}
  {\bibfnamefont {D.}~\bibnamefont {Gu\'ery-Odelin}}, \ and\ \bibinfo {author}
  {\bibfnamefont {J.~G.}\ \bibnamefont {Muga}},\ }\href {\doibase
  10.1103/PhysRevLett.104.063002} {\bibfield  {journal} {\bibinfo  {journal}
  {Phys. Rev. Lett.}\ }\textbf {\bibinfo {volume} {104}},\ \bibinfo {pages}
  {063002} (\bibinfo {year} {2010})}\BibitemShut {NoStop}%
\bibitem [{\citenamefont {del Campo}\ \emph {et~al.}(2012)\citenamefont {del
  Campo}, \citenamefont {Rams},\ and\ \citenamefont {Zurek}}]{delCampo2012}%
  \BibitemOpen
  \bibfield  {author} {\bibinfo {author} {\bibfnamefont {A.}~\bibnamefont {del
  Campo}}, \bibinfo {author} {\bibfnamefont {M.~M.}\ \bibnamefont {Rams}}, \
  and\ \bibinfo {author} {\bibfnamefont {W.~H.}\ \bibnamefont {Zurek}},\ }\href
  {\doibase 10.1103/PhysRevLett.109.115703} {\bibfield  {journal} {\bibinfo
  {journal} {Phys. Rev. Lett.}\ }\textbf {\bibinfo {volume} {109}},\ \bibinfo
  {pages} {115703} (\bibinfo {year} {2012})}\BibitemShut {NoStop}%
\bibitem [{\citenamefont {del Campo}(2013)}]{delCampo2013}%
  \BibitemOpen
  \bibfield  {author} {\bibinfo {author} {\bibfnamefont {A.}~\bibnamefont {del
  Campo}},\ }\href {\doibase 10.1103/PhysRevLett.111.100502} {\bibfield
  {journal} {\bibinfo  {journal} {Phys. Rev. Lett.}\ }\textbf {\bibinfo
  {volume} {111}},\ \bibinfo {pages} {100502} (\bibinfo {year}
  {2013})}\BibitemShut {NoStop}%
\bibitem [{\citenamefont {Torrontegui}\ \emph {et~al.}(2013)\citenamefont
  {Torrontegui}, \citenamefont {Ib{\'a\~n}ez}, \citenamefont
  {Mart{\'i}nez-Garaot}, \citenamefont {Modugno}, \citenamefont {del Campo},
  \citenamefont {Gu{\'e}ry-Odelin}, \citenamefont {Ruschhaupt}, \citenamefont
  {Chen},\ and\ \citenamefont {Muga}}]{Torrontegui2013}%
  \BibitemOpen
  \bibfield  {author} {\bibinfo {author} {\bibfnamefont {E.}~\bibnamefont
  {Torrontegui}}, \bibinfo {author} {\bibfnamefont {S.}~\bibnamefont
  {Ib{\'a\~n}ez}}, \bibinfo {author} {\bibfnamefont {S.}~\bibnamefont
  {Mart{\'i}nez-Garaot}}, \bibinfo {author} {\bibfnamefont {M.}~\bibnamefont
  {Modugno}}, \bibinfo {author} {\bibfnamefont {A.}~\bibnamefont {del Campo}},
  \bibinfo {author} {\bibfnamefont {D.}~\bibnamefont {Gu{\'e}ry-Odelin}},
  \bibinfo {author} {\bibfnamefont {A.}~\bibnamefont {Ruschhaupt}}, \bibinfo
  {author} {\bibfnamefont {X.}~\bibnamefont {Chen}}, \ and\ \bibinfo {author}
  {\bibfnamefont {J.~G.}\ \bibnamefont {Muga}},\ }\href {\doibase
  http://dx.doi.org/10.1016/B978-0-12-408090-4.00002-5} {\bibfield  {journal}
  {\bibinfo  {journal} {Adv. At. Mol. Opt. Phys.}\ }\textbf {\bibinfo {volume}
  {62}},\ \bibinfo {pages} {117 } (\bibinfo {year} {2013})}\BibitemShut
  {NoStop}%
\bibitem [{\citenamefont {Deng}\ \emph {et~al.}(2013)\citenamefont {Deng},
  \citenamefont {Wang}, \citenamefont {Liu}, \citenamefont {H\"anggi},\ and\
  \citenamefont {Gong}}]{Deng2013}%
  \BibitemOpen
  \bibfield  {author} {\bibinfo {author} {\bibfnamefont {J.}~\bibnamefont
  {Deng}}, \bibinfo {author} {\bibfnamefont {Q.-h.}\ \bibnamefont {Wang}},
  \bibinfo {author} {\bibfnamefont {Z.}~\bibnamefont {Liu}}, \bibinfo {author}
  {\bibfnamefont {P.}~\bibnamefont {H\"anggi}}, \ and\ \bibinfo {author}
  {\bibfnamefont {J.}~\bibnamefont {Gong}},\ }\href {\doibase
  10.1103/PhysRevE.88.062122} {\bibfield  {journal} {\bibinfo  {journal} {Phys.
  Rev. E}\ }\textbf {\bibinfo {volume} {88}},\ \bibinfo {pages} {062122}
  (\bibinfo {year} {2013})}\BibitemShut {NoStop}%
\bibitem [{\citenamefont {del Campo}\ \emph {et~al.}(2014)\citenamefont {del
  Campo}, \citenamefont {Goold},\ and\ \citenamefont
  {Paternostro}}]{delCampo2014}%
  \BibitemOpen
  \bibfield  {author} {\bibinfo {author} {\bibfnamefont {A.}~\bibnamefont {del
  Campo}}, \bibinfo {author} {\bibfnamefont {J.}~\bibnamefont {Goold}}, \ and\
  \bibinfo {author} {\bibfnamefont {M.}~\bibnamefont {Paternostro}},\ }\href
  {\doibase 10.1038/srep06208} {\bibfield  {journal} {\bibinfo  {journal} {Sci.
  Rep.}\ }\textbf {\bibinfo {volume} {4}},\ \bibinfo {pages} {6208} (\bibinfo
  {year} {2014})}\BibitemShut {NoStop}%
\bibitem [{\citenamefont {Beau}\ \emph {et~al.}(2016)\citenamefont {Beau},
  \citenamefont {Jaramillo},\ and\ \citenamefont {del Campo}}]{Beau2016}%
  \BibitemOpen
  \bibfield  {author} {\bibinfo {author} {\bibfnamefont {M.}~\bibnamefont
  {Beau}}, \bibinfo {author} {\bibfnamefont {J.}~\bibnamefont {Jaramillo}}, \
  and\ \bibinfo {author} {\bibfnamefont {A.}~\bibnamefont {del Campo}},\ }\href
  {\doibase 10.3390/e18050168} {\bibfield  {journal} {\bibinfo  {journal}
  {Entropy}\ }\textbf {\bibinfo {volume} {18}},\ \bibinfo {pages} {168}
  (\bibinfo {year} {2016})}\BibitemShut {NoStop}%
\bibitem [{\citenamefont {Juli\'a-D\'{\i}az}\ \emph {et~al.}(2012)\citenamefont
  {Juli\'a-D\'{\i}az}, \citenamefont {Torrontegui}, \citenamefont {Martorell},
  \citenamefont {Muga},\ and\ \citenamefont {Polls}}]{Julia-Diaz2012}%
  \BibitemOpen
  \bibfield  {author} {\bibinfo {author} {\bibfnamefont {B.}~\bibnamefont
  {Juli\'a-D\'{\i}az}}, \bibinfo {author} {\bibfnamefont {E.}~\bibnamefont
  {Torrontegui}}, \bibinfo {author} {\bibfnamefont {J.}~\bibnamefont
  {Martorell}}, \bibinfo {author} {\bibfnamefont {J.~G.}\ \bibnamefont {Muga}},
  \ and\ \bibinfo {author} {\bibfnamefont {A.}~\bibnamefont {Polls}},\ }\href
  {\doibase 10.1103/PhysRevA.86.063623} {\bibfield  {journal} {\bibinfo
  {journal} {Phys. Rev. A}\ }\textbf {\bibinfo {volume} {86}},\ \bibinfo
  {pages} {063623} (\bibinfo {year} {2012})}\BibitemShut {NoStop}%
\bibitem [{\citenamefont {Hatomura}(2018)}]{Hatomura2018}%
  \BibitemOpen
  \bibfield  {author} {\bibinfo {author} {\bibfnamefont {T.}~\bibnamefont
  {Hatomura}},\ }\href {http://stacks.iop.org/1367-2630/20/i=1/a=015010}
  {\bibfield  {journal} {\bibinfo  {journal} {New J. Phys.}\ }\textbf {\bibinfo
  {volume} {20}},\ \bibinfo {pages} {015010} (\bibinfo {year}
  {2018})}\BibitemShut {NoStop}%
\bibitem [{\citenamefont {Kyaw}\ and\ \citenamefont {Kwek}(2018)}]{Kyaw2018}%
  \BibitemOpen
  \bibfield  {author} {\bibinfo {author} {\bibfnamefont {T.~H.}\ \bibnamefont
  {Kyaw}}\ and\ \bibinfo {author} {\bibfnamefont {L.-C.}\ \bibnamefont
  {Kwek}},\ }\href {http://stacks.iop.org/1367-2630/20/i=4/a=045007} {\bibfield
   {journal} {\bibinfo  {journal} {New J. Phys.}\ }\textbf {\bibinfo {volume}
  {20}},\ \bibinfo {pages} {045007} (\bibinfo {year} {2018})}\BibitemShut
  {NoStop}%
\bibitem [{\citenamefont {Santos}\ and\ \citenamefont
  {Sarandy}(2015)}]{Santos2015}%
  \BibitemOpen
  \bibfield  {author} {\bibinfo {author} {\bibfnamefont {A.~C.}\ \bibnamefont
  {Santos}}\ and\ \bibinfo {author} {\bibfnamefont {M.~S.}\ \bibnamefont
  {Sarandy}},\ }\href {\doibase 10.1038/srep15775} {\bibfield  {journal}
  {\bibinfo  {journal} {Sci. Rep.}\ }\textbf {\bibinfo {volume} {5}},\ \bibinfo
  {pages} {15775} (\bibinfo {year} {2015})}\BibitemShut {NoStop}%
\bibitem [{\citenamefont {Santos}\ \emph {et~al.}(2016)\citenamefont {Santos},
  \citenamefont {Silva},\ and\ \citenamefont {Sarandy}}]{Santos2016}%
  \BibitemOpen
  \bibfield  {author} {\bibinfo {author} {\bibfnamefont {A.~C.}\ \bibnamefont
  {Santos}}, \bibinfo {author} {\bibfnamefont {R.~D.}\ \bibnamefont {Silva}}, \
  and\ \bibinfo {author} {\bibfnamefont {M.~S.}\ \bibnamefont {Sarandy}},\
  }\href {\doibase 10.1103/PhysRevA.93.012311} {\bibfield  {journal} {\bibinfo
  {journal} {Phys. Rev. A}\ }\textbf {\bibinfo {volume} {93}},\ \bibinfo
  {pages} {012311} (\bibinfo {year} {2016})}\BibitemShut {NoStop}%
\bibitem [{\citenamefont {Takahashi}(2017)}]{Takahashi2017}%
  \BibitemOpen
  \bibfield  {author} {\bibinfo {author} {\bibfnamefont {K.}~\bibnamefont
  {Takahashi}},\ }\href {\doibase 10.1103/PhysRevA.95.012309} {\bibfield
  {journal} {\bibinfo  {journal} {Phys. Rev. A}\ }\textbf {\bibinfo {volume}
  {95}},\ \bibinfo {pages} {012309} (\bibinfo {year} {2017})}\BibitemShut
  {NoStop}%
\bibitem [{\citenamefont {Jarzynski}(2013)}]{Jarzynski2013}%
  \BibitemOpen
  \bibfield  {author} {\bibinfo {author} {\bibfnamefont {C.}~\bibnamefont
  {Jarzynski}},\ }\href {\doibase 10.1103/PhysRevA.88.040101} {\bibfield
  {journal} {\bibinfo  {journal} {Phys. Rev. A}\ }\textbf {\bibinfo {volume}
  {88}},\ \bibinfo {pages} {040101} (\bibinfo {year} {2013})}\BibitemShut
  {NoStop}%
\bibitem [{\citenamefont {Deffner}\ \emph {et~al.}(2014)\citenamefont
  {Deffner}, \citenamefont {Jarzynski},\ and\ \citenamefont {del
  Campo}}]{Deffner2014}%
  \BibitemOpen
  \bibfield  {author} {\bibinfo {author} {\bibfnamefont {S.}~\bibnamefont
  {Deffner}}, \bibinfo {author} {\bibfnamefont {C.}~\bibnamefont {Jarzynski}},
  \ and\ \bibinfo {author} {\bibfnamefont {A.}~\bibnamefont {del Campo}},\
  }\href {\doibase 10.1103/PhysRevX.4.021013} {\bibfield  {journal} {\bibinfo
  {journal} {Phys. Rev. X}\ }\textbf {\bibinfo {volume} {4}},\ \bibinfo {pages}
  {021013} (\bibinfo {year} {2014})}\BibitemShut {NoStop}%
\bibitem [{\citenamefont {Okuyama}\ and\ \citenamefont
  {Takahashi}(2017)}]{Okuyama2017}%
  \BibitemOpen
  \bibfield  {author} {\bibinfo {author} {\bibfnamefont {M.}~\bibnamefont
  {Okuyama}}\ and\ \bibinfo {author} {\bibfnamefont {K.}~\bibnamefont
  {Takahashi}},\ }\href {\doibase 10.7566/JPSJ.86.043002} {\bibfield  {journal}
  {\bibinfo  {journal} {J. Phys. Soc. Jpn.}\ }\textbf {\bibinfo {volume}
  {86}},\ \bibinfo {pages} {043002} (\bibinfo {year} {2017})}\BibitemShut
  {NoStop}%
\bibitem [{\citenamefont {Sels}\ and\ \citenamefont
  {Polkovnikov}(2017)}]{Sels2017}%
  \BibitemOpen
  \bibfield  {author} {\bibinfo {author} {\bibfnamefont {D.}~\bibnamefont
  {Sels}}\ and\ \bibinfo {author} {\bibfnamefont {A.}~\bibnamefont
  {Polkovnikov}},\ }\href {\doibase 10.1073/pnas.1619826114} {\bibfield
  {journal} {\bibinfo  {journal} {Proc. Natl. Acad. Sci. USA}\ }\textbf
  {\bibinfo {volume} {114}},\ \bibinfo {pages} {E3909} (\bibinfo {year}
  {2017})}\BibitemShut {NoStop}%
\bibitem [{Note1()}]{Note1}%
  \BibitemOpen
  \bibinfo {note} {We can construct exact CD terms if all possible operators
  are taken into account for trial CD terms. However, it is usually difficult
  for many-body systems.}\BibitemShut {Stop}%
\bibitem [{\citenamefont {van Hemmen}\ and\ \citenamefont
  {S\"ut\"o}(1986)}]{vanHemmen1986}%
  \BibitemOpen
  \bibfield  {author} {\bibinfo {author} {\bibfnamefont {J.~L.}\ \bibnamefont
  {van Hemmen}}\ and\ \bibinfo {author} {\bibfnamefont {A.}~\bibnamefont
  {S\"ut\"o}},\ }\href {http://stacks.iop.org/0295-5075/1/i=10/a=001}
  {\bibfield  {journal} {\bibinfo  {journal} {Europhys. Lett.}\ }\textbf
  {\bibinfo {volume} {1}},\ \bibinfo {pages} {481} (\bibinfo {year}
  {1986})}\BibitemShut {NoStop}%
\bibitem [{\citenamefont {Hatomura}(2017)}]{Hatomura2017}%
  \BibitemOpen
  \bibfield  {author} {\bibinfo {author} {\bibfnamefont {T.}~\bibnamefont
  {Hatomura}},\ }\href {\doibase 10.7566/JPSJ.86.094002} {\bibfield  {journal}
  {\bibinfo  {journal} {J. Phys. Soc. Jpn.}\ }\textbf {\bibinfo {volume}
  {86}},\ \bibinfo {pages} {094002} (\bibinfo {year} {2017})}\BibitemShut
  {NoStop}%
\bibitem [{\citenamefont {Tanaka}\ and\ \citenamefont
  {Horiguchi}(1997)}]{Tanaka-Horiguchi-Jpn}%
  \BibitemOpen
  \bibfield  {author} {\bibinfo {author} {\bibfnamefont {K.}~\bibnamefont
  {Tanaka}}\ and\ \bibinfo {author} {\bibfnamefont {T.}~\bibnamefont
  {Horiguchi}},\ }\href
  {https://search.ieice.org/bin/summary.php?id=j80-a_12_2117} {\bibfield
  {journal} {\bibinfo  {journal} {Denshi Joho Tsushin Gakkai Ronbunshi A}\
  }\textbf {\bibinfo {volume} {80}},\ \bibinfo {pages} {2117} (\bibinfo {year}
  {1997})},\ \bibinfo {note} {in {Japanese}}\BibitemShut {NoStop}%
\bibitem [{\citenamefont {Tanaka}\ and\ \citenamefont
  {Horiguchi}(2000)}]{Tanaka2000}%
  \BibitemOpen
  \bibfield  {author} {\bibinfo {author} {\bibfnamefont {K.}~\bibnamefont
  {Tanaka}}\ and\ \bibinfo {author} {\bibfnamefont {T.}~\bibnamefont
  {Horiguchi}},\ }\href {\doibase
  10.1002/(SICI)1520-6440(200003)83:3<84::AID-ECJC9>3.0.CO;2-N} {\bibfield
  {journal} {\bibinfo  {journal} {Electro. Commun. Jpn. Part III}\ }\textbf
  {\bibinfo {volume} {83}},\ \bibinfo {pages} {84} (\bibinfo {year} {2000})},\
  \bibinfo {note} {{English} translation}\BibitemShut {NoStop}%
\bibitem [{\citenamefont {Brooke}\ \emph {et~al.}(1999)\citenamefont {Brooke},
  \citenamefont {Bitko}, \citenamefont {Rosenbaum},\ and\ \citenamefont
  {Aeppli}}]{Brooke1999}%
  \BibitemOpen
  \bibfield  {author} {\bibinfo {author} {\bibfnamefont {J.}~\bibnamefont
  {Brooke}}, \bibinfo {author} {\bibfnamefont {D.}~\bibnamefont {Bitko}},
  \bibinfo {author} {\bibfnamefont {T.~F.}\ \bibnamefont {Rosenbaum}}, \ and\
  \bibinfo {author} {\bibfnamefont {G.}~\bibnamefont {Aeppli}},\ }\href
  {\doibase 10.1126/science.284.5415.779} {\bibfield  {journal} {\bibinfo
  {journal} {Science}\ }\textbf {\bibinfo {volume} {284}},\ \bibinfo {pages}
  {779} (\bibinfo {year} {1999})}\BibitemShut {NoStop}%
\bibitem [{\citenamefont {Santoro}\ \emph {et~al.}(2002)\citenamefont
  {Santoro}, \citenamefont {Marto{\v n}{\'a}k}, \citenamefont {Tosatti},\ and\
  \citenamefont {Car}}]{Santoro2427}%
  \BibitemOpen
  \bibfield  {author} {\bibinfo {author} {\bibfnamefont {G.~E.}\ \bibnamefont
  {Santoro}}, \bibinfo {author} {\bibfnamefont {R.}~\bibnamefont {Marto{\v
  n}{\'a}k}}, \bibinfo {author} {\bibfnamefont {E.}~\bibnamefont {Tosatti}}, \
  and\ \bibinfo {author} {\bibfnamefont {R.}~\bibnamefont {Car}},\ }\href
  {\doibase 10.1126/science.1068774} {\bibfield  {journal} {\bibinfo  {journal}
  {Science}\ }\textbf {\bibinfo {volume} {295}},\ \bibinfo {pages} {2427}
  (\bibinfo {year} {2002})}\BibitemShut {NoStop}%
\bibitem [{\citenamefont {Marto{\v{n}}{\'a}k}\ \emph
  {et~al.}(2002)\citenamefont {Marto{\v{n}}{\'a}k}, \citenamefont {Santoro},\
  and\ \citenamefont {Tosatti}}]{Martonak2002}%
  \BibitemOpen
  \bibfield  {author} {\bibinfo {author} {\bibfnamefont {R.}~\bibnamefont
  {Marto{\v{n}}{\'a}k}}, \bibinfo {author} {\bibfnamefont {G.~E.}\ \bibnamefont
  {Santoro}}, \ and\ \bibinfo {author} {\bibfnamefont {E.}~\bibnamefont
  {Tosatti}},\ }\href {\doibase 10.1103/PhysRevB.66.094203} {\bibfield
  {journal} {\bibinfo  {journal} {Phys. Rev. B}\ }\textbf {\bibinfo {volume}
  {66}},\ \bibinfo {pages} {094203} (\bibinfo {year} {2002})}\BibitemShut
  {NoStop}%
\bibitem [{\citenamefont {Johnson}\ \emph {et~al.}(2011)\citenamefont
  {Johnson}, \citenamefont {Amin}, \citenamefont {Gildert}, \citenamefont
  {Lanting}, \citenamefont {Hamze}, \citenamefont {Dickson}, \citenamefont
  {Harris}, \citenamefont {Berkley}, \citenamefont {Johansson}, \citenamefont
  {Bunyk}, \citenamefont {Chapple}, \citenamefont {Enderud}, \citenamefont
  {Hilton}, \citenamefont {Karimi}, \citenamefont {Ladizinsky}, \citenamefont
  {Ladizinsky}, \citenamefont {Oh}, \citenamefont {Perminov}, \citenamefont
  {Rich}, \citenamefont {Thom}, \citenamefont {Tolkacheva}, \citenamefont
  {Truncik}, \citenamefont {Uchaikin}, \citenamefont {Wang}, \citenamefont
  {Wilson},\ and\ \citenamefont {Rose}}]{Johnson2011}%
  \BibitemOpen
  \bibfield  {author} {\bibinfo {author} {\bibfnamefont {M.~W.}\ \bibnamefont
  {Johnson}}, \bibinfo {author} {\bibfnamefont {M.~H.}\ \bibnamefont {Amin}},
  \bibinfo {author} {\bibfnamefont {S.}~\bibnamefont {Gildert}}, \bibinfo
  {author} {\bibfnamefont {T.}~\bibnamefont {Lanting}}, \bibinfo {author}
  {\bibfnamefont {F.}~\bibnamefont {Hamze}}, \bibinfo {author} {\bibfnamefont
  {N.}~\bibnamefont {Dickson}}, \bibinfo {author} {\bibfnamefont
  {R.}~\bibnamefont {Harris}}, \bibinfo {author} {\bibfnamefont {A.~J.}\
  \bibnamefont {Berkley}}, \bibinfo {author} {\bibfnamefont {J.}~\bibnamefont
  {Johansson}}, \bibinfo {author} {\bibfnamefont {P.}~\bibnamefont {Bunyk}},
  \bibinfo {author} {\bibfnamefont {E.~M.}\ \bibnamefont {Chapple}}, \bibinfo
  {author} {\bibfnamefont {C.}~\bibnamefont {Enderud}}, \bibinfo {author}
  {\bibfnamefont {J.~P.}\ \bibnamefont {Hilton}}, \bibinfo {author}
  {\bibfnamefont {K.}~\bibnamefont {Karimi}}, \bibinfo {author} {\bibfnamefont
  {E.}~\bibnamefont {Ladizinsky}}, \bibinfo {author} {\bibfnamefont
  {N.}~\bibnamefont {Ladizinsky}}, \bibinfo {author} {\bibfnamefont
  {T.}~\bibnamefont {Oh}}, \bibinfo {author} {\bibfnamefont {I.}~\bibnamefont
  {Perminov}}, \bibinfo {author} {\bibfnamefont {C.}~\bibnamefont {Rich}},
  \bibinfo {author} {\bibfnamefont {M.~C.}\ \bibnamefont {Thom}}, \bibinfo
  {author} {\bibfnamefont {E.}~\bibnamefont {Tolkacheva}}, \bibinfo {author}
  {\bibfnamefont {C.~J.~S.}\ \bibnamefont {Truncik}}, \bibinfo {author}
  {\bibfnamefont {S.}~\bibnamefont {Uchaikin}}, \bibinfo {author}
  {\bibfnamefont {J.}~\bibnamefont {Wang}}, \bibinfo {author} {\bibfnamefont
  {B.}~\bibnamefont {Wilson}}, \ and\ \bibinfo {author} {\bibfnamefont
  {G.}~\bibnamefont {Rose}},\ }\href {\doibase 10.1038/nature10012} {\bibfield
  {journal} {\bibinfo  {journal} {Nature}\ }\textbf {\bibinfo {volume} {473}},\
  \bibinfo {pages} {194} (\bibinfo {year} {2011})}\BibitemShut {NoStop}%
\bibitem [{\citenamefont {Smolin}\ and\ \citenamefont
  {Smith}(2014)}]{Smolin2014}%
  \BibitemOpen
  \bibfield  {author} {\bibinfo {author} {\bibfnamefont {J.~A.}\ \bibnamefont
  {Smolin}}\ and\ \bibinfo {author} {\bibfnamefont {G.}~\bibnamefont {Smith}},\
  }\href {\doibase 10.3389/fphy.2014.00052} {\bibfield  {journal} {\bibinfo
  {journal} {Front. Phys.}\ }\textbf {\bibinfo {volume} {2}},\ \bibinfo {pages}
  {52} (\bibinfo {year} {2014})}\BibitemShut {NoStop}%
\bibitem [{\citenamefont {Wang}\ \emph {et~al.}(2013)\citenamefont {Wang},
  \citenamefont {R{\o}nnow}, \citenamefont {Boixo}, \citenamefont {Isakov},
  \citenamefont {Wang}, \citenamefont {Wecker}, \citenamefont {Lidar},
  \citenamefont {Martinis},\ and\ \citenamefont {Troyer}}]{Wang2013}%
  \BibitemOpen
  \bibfield  {author} {\bibinfo {author} {\bibfnamefont {L.}~\bibnamefont
  {Wang}}, \bibinfo {author} {\bibfnamefont {T.~F.}\ \bibnamefont {R{\o}nnow}},
  \bibinfo {author} {\bibfnamefont {S.}~\bibnamefont {Boixo}}, \bibinfo
  {author} {\bibfnamefont {S.~V.}\ \bibnamefont {Isakov}}, \bibinfo {author}
  {\bibfnamefont {Z.}~\bibnamefont {Wang}}, \bibinfo {author} {\bibfnamefont
  {D.}~\bibnamefont {Wecker}}, \bibinfo {author} {\bibfnamefont {D.~A.}\
  \bibnamefont {Lidar}}, \bibinfo {author} {\bibfnamefont {J.~M.}\ \bibnamefont
  {Martinis}}, \ and\ \bibinfo {author} {\bibfnamefont {M.}~\bibnamefont
  {Troyer}},\ }\href {https://arxiv.org/abs/1305.5837} {\bibfield  {journal}
  {\bibinfo  {journal} {arXiv preprint arXiv:1305.5837}\ } (\bibinfo {year}
  {2013})}\BibitemShut {NoStop}%
\bibitem [{\citenamefont {Boixo}\ \emph {et~al.}(2014)\citenamefont {Boixo},
  \citenamefont {R{\o}nnow}, \citenamefont {Isakov}, \citenamefont {Wang},
  \citenamefont {Wecker}, \citenamefont {Lidar}, \citenamefont {Martinis},\
  and\ \citenamefont {Troyer}}]{Boixo2014}%
  \BibitemOpen
  \bibfield  {author} {\bibinfo {author} {\bibfnamefont {S.}~\bibnamefont
  {Boixo}}, \bibinfo {author} {\bibfnamefont {T.~F.}\ \bibnamefont
  {R{\o}nnow}}, \bibinfo {author} {\bibfnamefont {S.~V.}\ \bibnamefont
  {Isakov}}, \bibinfo {author} {\bibfnamefont {Z.}~\bibnamefont {Wang}},
  \bibinfo {author} {\bibfnamefont {D.}~\bibnamefont {Wecker}}, \bibinfo
  {author} {\bibfnamefont {D.~A.}\ \bibnamefont {Lidar}}, \bibinfo {author}
  {\bibfnamefont {J.~M.}\ \bibnamefont {Martinis}}, \ and\ \bibinfo {author}
  {\bibfnamefont {M.}~\bibnamefont {Troyer}},\ }\href {\doibase
  10.1038/nphys2900} {\bibfield  {journal} {\bibinfo  {journal} {Nature
  Physics}\ }\textbf {\bibinfo {volume} {10}},\ \bibinfo {pages} {218}
  (\bibinfo {year} {2014})}\BibitemShut {NoStop}%
\bibitem [{\citenamefont {Shin}\ \emph {et~al.}(2014)\citenamefont {Shin},
  \citenamefont {Smith}, \citenamefont {Smolin},\ and\ \citenamefont
  {Vazirani}}]{Shin2014}%
  \BibitemOpen
  \bibfield  {author} {\bibinfo {author} {\bibfnamefont {S.~W.}\ \bibnamefont
  {Shin}}, \bibinfo {author} {\bibfnamefont {G.}~\bibnamefont {Smith}},
  \bibinfo {author} {\bibfnamefont {J.~A.}\ \bibnamefont {Smolin}}, \ and\
  \bibinfo {author} {\bibfnamefont {U.}~\bibnamefont {Vazirani}},\ }\href
  {https://arxiv.org/abs/1401.7087} {\bibfield  {journal} {\bibinfo  {journal}
  {arXiv preprint arXiv:1401.7087}\ } (\bibinfo {year} {2014})}\BibitemShut
  {NoStop}%
\bibitem [{\citenamefont {Albash}\ \emph {et~al.}(2015)\citenamefont {Albash},
  \citenamefont {R{\o}nnow}, \citenamefont {Troyer},\ and\ \citenamefont
  {Lidar}}]{Albash2015reexamining}%
  \BibitemOpen
  \bibfield  {author} {\bibinfo {author} {\bibfnamefont {T.}~\bibnamefont
  {Albash}}, \bibinfo {author} {\bibfnamefont {T.~F.}\ \bibnamefont
  {R{\o}nnow}}, \bibinfo {author} {\bibfnamefont {M.}~\bibnamefont {Troyer}}, \
  and\ \bibinfo {author} {\bibfnamefont {D.~A.}\ \bibnamefont {Lidar}},\ }\href
  {\doibase 10.1140/epjst/e2015-02346-0} {\bibfield  {journal} {\bibinfo
  {journal} {Eur. Phys. J. Special Topics}\ }\textbf {\bibinfo {volume}
  {224}},\ \bibinfo {pages} {111} (\bibinfo {year} {2015})}\BibitemShut
  {NoStop}%
\bibitem [{\citenamefont {Rams}\ \emph {et~al.}(2016)\citenamefont {Rams},
  \citenamefont {Mohseni},\ and\ \citenamefont {del Campo}}]{Rams2016}%
  \BibitemOpen
  \bibfield  {author} {\bibinfo {author} {\bibfnamefont {M.~M.}\ \bibnamefont
  {Rams}}, \bibinfo {author} {\bibfnamefont {M.}~\bibnamefont {Mohseni}}, \
  and\ \bibinfo {author} {\bibfnamefont {A.}~\bibnamefont {del Campo}},\ }\href
  {http://stacks.iop.org/1367-2630/18/i=12/a=123034} {\bibfield  {journal}
  {\bibinfo  {journal} {New J. Phys.}\ }\textbf {\bibinfo {volume} {18}},\
  \bibinfo {pages} {123034} (\bibinfo {year} {2016})}\BibitemShut {NoStop}%
\bibitem [{\citenamefont {Susa}\ \emph {et~al.}(2018)\citenamefont {Susa},
  \citenamefont {Yamashiro}, \citenamefont {Yamamoto},\ and\ \citenamefont
  {Nishimori}}]{Susa2018}%
  \BibitemOpen
  \bibfield  {author} {\bibinfo {author} {\bibfnamefont {Y.}~\bibnamefont
  {Susa}}, \bibinfo {author} {\bibfnamefont {Y.}~\bibnamefont {Yamashiro}},
  \bibinfo {author} {\bibfnamefont {M.}~\bibnamefont {Yamamoto}}, \ and\
  \bibinfo {author} {\bibfnamefont {H.}~\bibnamefont {Nishimori}},\ }\href
  {\doibase 10.7566/JPSJ.87.023002} {\bibfield  {journal} {\bibinfo  {journal}
  {J. Phys. Soc. Jpn.}\ }\textbf {\bibinfo {volume} {87}},\ \bibinfo {pages}
  {023002} (\bibinfo {year} {2018})}\BibitemShut {NoStop}%
\bibitem [{\citenamefont {Rieger}(1998)}]{Rieger1998}%
  \BibitemOpen
  \bibfield  {author} {\bibinfo {author} {\bibfnamefont {H.}~\bibnamefont
  {Rieger}},\ }in\ \href@noop {} {\emph {\bibinfo {booktitle} {Advances in
  Computer Simulation}}},\ \bibinfo {editor} {edited by\ \bibinfo {editor}
  {\bibfnamefont {I.}~\bibnamefont {Kert{\'e}sz}, \bibfnamefont
  {J{\'a}nosand~Kondor}}}\ (\bibinfo  {publisher} {Springer Berlin
  Heidelberg},\ \bibinfo {address} {Berlin, Heidelberg},\ \bibinfo {year}
  {1998})\ pp.\ \bibinfo {pages} {122--158}\BibitemShut {NoStop}%
\end{thebibliography}%

\end{document}